\documentclass[prd,aps,showpacs,showkeys]{revtex4-1}
\usepackage{bm,amsfonts,latexsym,amsmath,amssymb,amsbsy,graphicx}

\newcommand{\bb}{\begin{eqnarray}}
\newcommand{\ee}{\end{eqnarray}}
\newcommand{\ba}{\begin{align}}
\newcommand{\ea}{\end{align}}

\begin{document}

\title{\bf Vacuum polarization of charged massless fermions in Coulomb and Aharonov--Bohm fields}
\author{V.R. Khalilov \footnote{Corresponding author}}\email{khalilov@phys.msu.ru}
\affiliation{Faculty of Physics, M.V. Lomonosov Moscow State University, 119991,
Moscow, Russia}
\author{I.V. Mamsurov}
\affiliation{Faculty of Physics, M.V. Lomonosov Moscow State University, 119991,
Moscow, Russia}

\begin{abstract}
Vacuum polarization of charged massless fermions is investigated in
the superposition of Coulomb and Aharonov--Bohm (AB) potentials in 2+1 dimensions.
For this purpose we construct the Green function of the two-dimensional Dirac equation
with Coulomb and AB potentials  (via the regular and
irregular solutions of the  radial Dirac equation) and
calculate the vacuum  polarization charge density  in these fields
in the so-called subcritical and supercritical regimes.
 The role of the self-adjoint extension parameter is discussed
 in terms of the physics of problem.
We hope that our results will be helpful in  the more deep understanding
the fundamental problem of quantum electrodynamics and can be applied to the
problems of charged impurity screening in graphene with taking into consideration the electron spin.
\end{abstract}



\maketitle

%\pacs{03.65.-w, 03.65.Pm, 03.65.Ge, 04.20.Jb}

\section{Introduction}

%PACS numbers: 73.22.Pr, 81.05.ue, 03.65.Pm

The vacuum instability  in the supercritical Coulomb field is one of the important
problem of quantum electrodynamics that is exhaustively
studied  in \cite{001,002,003,004,005,006,007,grrein}.
Main physical quantity  related to this problem is the induced vacuum polarization \cite{004,007,grrein,01,02,03,04}.
New great interest to the vacuum instability in the supercritical
Coulomb field was revived in connection with the Coulomb impurity problem in
 graphene. The effective fine structure
constant in graphene is large, which gives the new possibility
to study the two-dimensional quantum electrodynamics in the strong-coupling
(in fact, supercritical Coulomb) regime and the existence of
charged Fermi quasiparticles in graphene makes  experimentally
feasible to observe the vacuum polarization in strong Coulomb field.

From the physical point of view there are two (subcritical and supercritical)
regimes, depending mainly on the magnitude of Coulomb field
charge. Theoretically, charged impurity screening in graphene
in terms of vacuum polarization  were investigated  in  \cite{mik,vp11a,as11b,as112,bsaso,mmfdsn,kn11c,12,vnkbu,vnkvmp,vmpvnk,13a} and
in  comprehensive reviews \cite{7,review}. These studies have shown that
the vacuum polarization charge density  is localized at
the potential center \cite{vp11a,bsaso,12} in the subcritical Coulomb potential
while the vacuum charge density induced by the supercritical Coulomb potential has
the form $c/r^2$ \cite{vp11a,as11b}  causing  a modification of Coulomb law
at large distances. This behavior could be expected on dimensional
grounds: $\delta(r)$ and $c/r^2$ are the only dimensionally
consistent possibilities due to the scaling invariance of the massless Dirac equation
(in the absence of  any intrinsic length scale) \cite{review}.
It will be remembered that close to the so-called Dirac points,
charged quasiparticle excitations in the potential of graphene lattice
are massless Dirac-like fermions characterized by a linear dispersion
relation \cite{7,review,6} and so a single electron dynamics in graphene
is described by a massless two-component Dirac equation \cite{vp11a,as11b,11,13a,7,review,ksn}.

New important results related to screening in graphene
were obtained in \cite{as11b,12} with taking into account
electron-electron interactions in a self-consistent
renormalization group treatment.
It turns out to be the system self-consistently rearranges
itself so that electrons at large distances never
feel a supercritical effective coupling and the subcritical
(stable) situation is therefore protected. This conclusion
agrees with expectations for the corresponding problem
in the convenient quantum electrodynamics, where
the vacuum polarization charge in super-heavy
nuclei behaves in such a way as to reduce the supercritical charge
of nucleus to the threshold value \cite{muraf} (see also \cite{terkh},
where the problem was investigated for super-heavy
nuclei in the presence of a superstrong constant uniform magnetic field).

The  wonderful quantum phenomenon was revealed in \cite{15}: the induced
current density in graphene in the field of a solenoid turns out to be a
finite periodic function of the magnetic flux.
The  induced polarization current in the QED$_{2+1}$ with an Aharonov--Bohm  potential
for  massive and massless charged fermions was studied in \cite{kh1}.
The induced electric current  due to vacuum polarization in the AB potential
was observed in \cite{34a} in ``a quantum-tunneling system using two-dimensional ionic
structures in a linear Paul trap''.

The dynamics of charged fermions in the superposition of
Coulomb and AB potentials  is governed by a singular Dirac Hamiltonian that requires
the supplementary definition in order for it to be treated
as a self-adjoint quantum-mechanical operator. So we need
to determine the self-adjoint Dirac Hamiltonians and then to construct the correct Green function
of the Dirac equation in the superposition of Coulomb and AB potentials.
In such a superposition the subcritical and supercritical
regimes are determined by the magnitudes of parameters
(as well as the relations between them) characterizing the Coulomb
and  AB potentials.

A main feature of the supercritical and (at some magnitudes of parameters) subcritical regimes
is a nonuniqueness of the self-adjoint Dirac Hamiltonian; there exists a one-parameter family of
self-adjoint Dirac Hamiltonians specified by additional boundary conditions
at the origin \cite{17}. This is a manifestation of a nontrivial
physics inside the origin and an interpretation of
self-adjoint extension parameters is a purely physical problem \cite{17}.
  For example,  in an AB field the magnetic flux within the interior of the vortex
  determines the effective Hamiltonian outside it; the extensions can be parameterized  by nontrivial boundary
conditions on the wave functions at the origin and different choices lead
to inequivalent physical cases \cite{phg}. We can determine the self-adjoint
extension parameter in terms of the parameter $R$ that is the
finite radius of a real solenoid \cite{kh1}. The self-adjoint extension
method was used  to determine
bound states of massive fermions in the Aharonov--Bohm-like fields \cite{kh2} and
of a magnetic dipole moment in electric and magnetic fields generated
by an infinitely long charged solenoid, carrying a magnetic field \cite{sil}.

Here we study the vacuum polarization of charged massless fermions
in the superposition of Coulomb and AB potentials in 2+1 dimensions.
We calculate  the induced charge density in the vacuum in the subcritical
and supercritical regimes, for the first time, using the Green function
of the two-dimensional Dirac equation with Coulomb and AB potentials.
 The presence of AB potential in our model gives
the possibility  to estimate effects, which are due to the
interaction of the electron spin magnetic moment
and the Aharonov--Bohm magnetic field.
Since the interaction potential is repulsive or attractive
for different signs of spin projection this feature  must be taken into account
in the behavior of wave functions at the origin.

We shall adopt the units where $c=\hbar=1$.

\section{Induced vacuum charge density}

\subsection{Solutions and Green function of the Dirac Hamiltonian}

We remind that the Dirac $\gamma^{\mu}$-matrix algebra is known to be represented in terms of the two-dimensional Pauli matrices  $\gamma^0= \sigma_3,\quad \gamma^1=is\sigma_1,\quad \gamma^2=i\sigma_2$ where  the parameter $s=\pm 1$ can be introduced to label two types of fermions in accordance with the
signature of the two-dimensional Dirac matrices \cite{27}; for the case of massive fermions it can be applied to characterize two states of the fermion spin (spin "up" and "down")  \cite{4}.

We also note that by Coulomb potential in 2+1 dimensions, we mean potential
that decreases as $1/r$ with the distance from the source, having in mind that in a physical situation (e.g., in graphene), although the electrons move in a plane, their Coulomb interaction with the external field of the pointlike charge of an impurity occurs in a physical (three-dimensional) space and the electric field strength of the impurity is a three-dimensional (not two-dimensional) vector. Therefore, the potential $A_0(r) \sim 1/r$ (and not $A_0(r) \sim \log r$, as would be
the case in 2+1 dimensions) does not satisfy the
two-dimensional Poisson equation with a pointlike source at the origin.
Besides, in real physical space, because of the existence of finite magnetic
flux inside solenoid $\Phi=2\pi B$ the singular term including the spin parameter appears
in the form of an additional delta-function interaction of spin with magnetic field of solenoid
${\bf H}=(0,\,0,\,H)=\nabla\times {\bf A}= B\pi\delta({\bf r})$
in the Dirac equation squared. The additional potential
$-seB\delta(r)/r$ will be taken into account by boundary conditions. It will be noted that
such kind of point interaction also appears in several Aharonov--Bohm-like problems \cite{asp1,sa1}.

The Dirac Hamiltonian for a fermion of the mass $m$ and charge
$e=-e_0<0$ in an   Aharonov--Bohm
$A_0=0$, $A_r=0$, $A_{\varphi}=B/r$, $r=\sqrt{x^2+y^2}$, $\varphi=\arctan(y/x)$
and Coulomb $A_0(r) =a/e_0r$, $A_r=0$, $A_{\varphi}=0$, $a>0$
potentials, is
\bb
 H_D=\sigma_1P_2-s\sigma_2P_1+\sigma_3 m-e_0A_0(r),\label{diham}
\ee
where $P_\mu = -i\partial_{\mu} - eA_{\mu}$ is the
generalized fermion momentum operator (a three-vector).
The Hamiltonian (\ref{diham}) should be defined as a self-adjoint operator in the Hilbert space
of square-integrable two-spinors $\Psi({\bf r})$.
The  total Dirac momentum operator  $J=-i\partial/\partial\varphi+ s\sigma_3/2$ commutes with  $H_D$.
Eigenfunctions of the Hamiltonian (\ref{diham}) are (see, \cite{khlee1})
\bb
 \Psi(t,{\bf r}) = \frac{1}{\sqrt{2\pi r}}
\left( \begin{array}{c}
f(r)\\
g(r)e^{is\varphi}
\end{array}\right)\exp(-iEt+il\varphi)~, \label{three}
\ee
where $E$ is  the fermion energy, $l$ is the integer quantum number.
The wave function $\Psi$ is an eigenfunction of the
operator $J$ with eigenvalue $j=\pm (l+s/2)$ in terms of the angular momentum $l$ and
\bb \check h F(r)= EF(r), \quad F(r)=\left(
\begin{array}{c}
f(r)\\
g(r)\end{array}\right), \label{radh}\ee
where
\bb
\check h=is\sigma_2\frac{d}{dr}+\sigma_1\frac{l+\mu+s/2}{r}+\sigma_3m-\frac{a}{r},\quad \mu\equiv e_0B. \label{radh0}
\ee
It will be noted that the massless fermions do not have spin degree of freedom in 2+1 dimensions \cite{jacknai},
nevertheless, the Dirac Hamiltonian (\ref{diham})  keeps the introduced spin parameter.

The induced current density due to to vacuum polarization is
determined by the three-vector $j_{\mu}({\bf r})$, which is
expressed in terms of the single-particle Green function of the Dirac equation as
\bb
 j_{\mu}({\bf r})=-\frac{e}{2}\int\limits_{C}\frac{dE}{2\pi i}{\rm tr}G({\bf r}, {\bf r'}; E)\gamma_{\mu},
\label{cur0}
\ee
where $C$ is the path in the complex plane of $E$ enclosing all the singularities
along the real axis $E$ depending upon the choice of the Fermi level $E_F$.
The Green function $G$ can be expanded in eigenfunctions of the operator $J$.
Since the induced charge density in the vacuum
is divergent and thus needs the renormalization it is helpful first to consider
the model with charged massive fermions.
For such a model, the radial parts (the doublets) of above eigenvalues must satisfy  the two-dimensional Dirac equation ({\ref{radh}).
Then the radial partial Green's function $G_l(r, r'; E)$ is given by (just as in 3+1 dimensions \cite{grrein})
\bb
G_l(r, r'; E)\gamma^0=\frac{1}{{\rm W}(E)}[\Theta(r'-r)U_R(r)U^{\dagger}_I(r')+
\Theta(r-r')U_I(r)U^{\dagger}_R(r')],
\label{green5}
\ee
where ${\rm W}(E)$ is the  ($r$-independent) Wronskian, defined by two doublets $V$ and $F$
as ${\rm Wr}(V, F)=Vi\sigma_2F=(v_1f_2-f_1v_2)$ and $U_R(r)$ and $U_I(r)$ are the regular and irregular solutions of the radial Dirac equation $(\check h-E)U(r)=0$; the regular (irregular) solutions are integrable at $r\to 0$ ($r\to\infty$).
We see that the problem is reduced to constructing the self-adjoint radial Hamiltonian $\check h$
 in the Hilbert space of  doublets $F(r)$ square-integrable on the half-line.

Since the initial radial Dirac operator  is not determined as an unique self-adjoint
operator the additional specification of its domain, given with the real parameter $\xi$ (the self-adjoint extension parameter)  is required in terms of the self-adjoint boundary conditions.
Any correct doublet $F(r)$   of the Hilbert space
 must satisfy the self-adjoint boundary condition \cite{17,khlee1,khlee}
\bb
 (F^{\dagger}(r)i\sigma_2 F(r))|_{r=0}= (\bar f_1f_2-\bar f_2f_1)|_{r=0} =0. \label{bouncon}
\ee
Physically, the self-adjoint boundary conditions  show that the probability current density  is equal to zero at the origin.

We shall apply as the solutions of the radial Dirac
equation (\ref{radh0}) the doublets found in \cite{20}
\bb
F_R = \left( \begin{array}{c} f_R(r, \gamma, E) \\ g_R(r, \gamma, E) \end{array} \right),
F_I = \left( \begin{array}{c} f_I(r, \gamma, E) \\ g_I(r, \gamma, E) \end{array} \right),
\label{WF1}
\ee
where
\bb
f_R(r, \gamma, E)=\frac{\sqrt{m+E}}{x} \left( A_R M_{aE/\lambda +s/2, \gamma}
(x) + C_R M_{aE/\lambda -s/2, \gamma} (x) \right), \phantom{mmmmmmmmmmm}\nonumber \\
g_R(r, \gamma, E)=\frac{\sqrt{m-E}}{x} \left( A_R M_{aE/\lambda +s/2, \gamma}
(x) - C_R M_{aE/\lambda -s/2, \gamma} (x) \right),\quad \frac{C_R}{A_R}=\frac{s\gamma -aE/\lambda}{\nu+ma/\lambda},
\label{base}
\ee
\bb
f_I(r, \gamma, E)=\frac{\sqrt{m+E}}{x} \left( A_I W_{aE/\lambda +s/2, \gamma}
(x) + C_I W_{aE/\lambda -s/2, \gamma} (x) \right),\phantom{mmmmmmmmmmm} \nonumber \\
g_I(r, \gamma, E)=\frac{\sqrt{m-E}}{x} \left( A_I W_{aE/\lambda +s/2, \gamma}
(x) - C_I W_{aE/\lambda -s/2, \gamma} (x) \right),\quad
\frac
{C_I}{A_I}=(ma/\lambda-s\nu)^s. \label{BASE}
\ee
Here
\bb
x=2\lambda r, \quad \lambda = \sqrt{m^2 - E^2}, \quad \gamma = \sqrt{\nu^2
-a^2}, \quad \nu=|l+\mu+s/2|, \label{Note}
\ee
$A_R, A_I, C_R, C_I$ are numerical coefficients and the Whittaker functions $M_{a,b}(x)$ and $W_{c,d}(x)$ represent the regular and irregular solutions.

For $a^2\leq \nu^2$ $\gamma$ is real, for $a^2>\nu^2$
$\gamma=i\sqrt{a^2-\nu^2}\equiv i\sigma$ is imaginary.
The quantities $q=\sqrt{\nu^2-\gamma^2}$ and  $q_c=\nu \Leftrightarrow\gamma=0$ are called the effective and   critical charge, respectively; it is helpful also to determine $q_u=\sqrt{\nu^2-1/4}\Leftrightarrow\gamma=1/2$.

\subsection{Induced charge density in the subcritical range}

In the subcritical range  for $q\leq q_u$, $\gamma\geq 1/2$, we can chose as the regular solutions  only  ones $F_R(r)$ vanishing at $r=0$; for $0<\gamma<1/2$ ($q_u<q<q_c$) the regular solutions $U_R(r)$ must satisfy the self-adjoint boundary condition (\ref{bouncon}) and should be chosen in the form of linear combination of the functions $F_R(r)$ and $F_I(r)$ \cite{17,khlee1}
\bb
U_R(r)=F_R(r)+\xi F_I(r).
\label{mainf}
\ee
The ($r$-independent) Wronskian is easily calculated to be
\bb
{\rm Wr}(F_R, F_I)\equiv {\rm W}(E, \gamma) = (g_R f_I - f_R g_I) = -2 A_R A_I \frac{\Gamma
(2\gamma)}{\Gamma (\gamma + 1/2 -s/2 -aE/\lambda)}
\frac{s\gamma}{\nu+ma/\lambda}
\label{wr1}
\ee
where $\Gamma(z)$ is the Gamma function \cite{GR} and, therefore, in the subcritical range
the single-particle Green function is completely determined.
One can show that the  contribution into the renormalized induced charge density coming from range
$0<\gamma<1/2$ is small for any $\xi$, therefore it is enough to consider the case $\xi=0$ in the subcritical range. Thus, we should chose as the regular solutions  the functions $F_R(r)$ for all $\gamma>0$ to obtain
\bb
{\rm tr} G_{\nu} ({\bf r}, {\bf r'}; E) \gamma^0 = \sum^{+1}_{s=-1}
\sum^{+\infty}_{l=-\infty}\frac{f_I f_R + g_I g_R}{2\pi s{\rm W}(E, \gamma)}.
\label{TRACE}
\ee
After some calculations, we obtain
\bb
{\rm tr} G_{\nu} ({\bf r}, {\bf r'}; E) \gamma^0 = -\frac{1}{2\pi \lambda^2 r^2}
\sum^{+\infty}_{l=-\infty} \frac{\Gamma(\gamma -
aE/\lambda)}{\Gamma(2\gamma +1)} \left[ (m^2 a/\lambda +E(x
-2aE/\lambda -1 )) M_{aE/\lambda +1/2, \gamma} (x) W_{aE/\lambda
+1/2, \gamma} (x) + \right. \nonumber \\
\left. + m^2a/\lambda (\gamma -aE/\lambda)M_{aE/\lambda -1/2,
\gamma} (x) W_{aE/\lambda -1/2, \gamma} (x) +
Ex\frac{d}{dx}(M_{aE/\lambda +1/2, \gamma} (x) W_{aE/\lambda +1/2,
\gamma} (x)) \right], \phantom{mmmm}\label{Calcul1}
\ee
where now $\gamma=\sqrt{\nu^2
-a^2}$, $\nu =l+\mu+1/2$.

We note that the singularities of $G_{\nu}({\bf r}, {\bf r'}; E)$ can be
simple poles associated with the discrete spectrum (in the range $-m<E<m$),
and two cuts $(-\infty,-m]$ and $[m,\infty)$ associated with the continuum
spectrum in the ranges $|E|\geq m$ \cite{20}.

  For the partial Green function in a Coulomb field in 3+1 dimensions, the path $C$ may be
deformed to run along the singularities on the real $E$ axis as follows:
$C=C_-+C_p+C_+$, where $C_-$ is the path along the negative
real $E$ axis (${\rm Re}E<0$) from $-\infty$ to $0$ turning
around at $E=0$ with positive
orientation, $C_p$ is a circle around the bound states singularities
with $-m< E<0$ (if we chose $E_F=-m$),
and $C_+$ is the path along the positive real $E$ axis (${\rm Re}E>0$)
from $\infty$ to $0$ but with negative orientation (i.e. clockwise path)
turning around at $E=0$ \cite{grrein}.
 In the considered case in 2+1 dimensions the path $C$ may be
deformed in the similar way \cite{kh1}.

One can show that the contour of integration $C$ with respect
to $E$ can be deformed to coincide with the imaginary axis and we obtain:
\bb
j_0({\bf r}) = -e \int\limits_{-\infty}^{+\infty}\frac{dE}{2\pi}{\rm tr}
G_{\nu} ({\bf r},{\bf r}^{\prime}, iE) \gamma^0. \label{Density}
\ee
Applying the following integral representation \cite{GR}
\bb
M_{aE/\lambda \pm 1/2,\gamma}(x) W_{aE/\lambda \pm 1/2,\gamma}(x)
=\frac{x\Gamma(2\gamma +1)}{\Gamma(1/2 +\gamma -aE/\lambda \mp 1/2)}
\int\limits_{0}^{\infty} e^{-x\cosh s}
[\coth(s/2)]^{2aE/\lambda \pm 1} I_{2\gamma}(x\sinh s) ds, \phantom{mmm}
\label{Form}
\ee
it is convenient  to represent the induced charge density in the form
\bb
j_0(r) = - \frac{2e}{\pi^2 r } \sum_{l=-\infty}^{+\infty}
\int\limits_{0}^{\infty} dE \int\limits_{0}^{\infty} dt
e^{-2\lambda r \coth t} \left( 2a \cos(2aE/\lambda) \coth t I_{2\gamma} (2\lambda
r/\sinh t) -\right. \nonumber\\ - \left. \frac{2Er}{\sinh t} \sin(2aE/\lambda)
I^{\prime}_{2\gamma} (2\lambda r/\sinh t) \right), \phantom{mmmmmm} \label{Result}
\ee
where now $\lambda =\sqrt{m^2+E^2}$, $I_{\mu}(z)$ is the modified Bessel function of the first kind  and
the prime (here and below) denotes the derivative of function with respect to argument.

Let us write  $\mu=[\mu]+\alpha\equiv n+\alpha$, where
$[\mu]\equiv n$ denotes the largest integer $\le \mu$, and $1>\alpha\ge 0$.
Hence $n=0, 1, 2, \ldots$ for $\mu>0$ and $n=-1, -2, -3, \ldots$ for $\mu<0$.
Since signs of $e$ and $B$ are fixed it is enough to consider the only
  case $\mu>0$.  Then, denoting  $\nu_{+} = l +\alpha +1/2$,
$\nu_{-} = l-\alpha +1/2$, $\gamma_{+} =\sqrt{\nu^{2}_{+} -a^2}$,
$\gamma_{-} =\sqrt{\nu^{2}_{-} -a^2}$, where here and in all formulas below $l\equiv l+n$,
we rewrite the induced charge density in the form
\bb
j_0(r) = - \frac{2e}{\pi^2 r } \sum_{l=0}^{+\infty}
\int\limits_{0}^{\infty} dE \int\limits_{0}^{\infty} dt
e^{-2\lambda r \coth t} \left( 2a \cos(2aE/\lambda) \coth t (I_{2\gamma_{+}}
(2\lambda r/\sinh t) + I_{2\gamma_{-}} (2\lambda
r/\sinh t)) -\right. \nonumber \\
- \left. \frac{2Er}{\sinh t} \sin(2aE/\lambda)
(I^{\prime}_{2\gamma_{+}} (2\lambda r/\sinh
t)+I^{\prime}_{2\gamma_{-}} (2\lambda r/\sinh t)) \right).
\label{Res1}
\ee
 emphasized the prime (here and below) denotes
We note that $j_0$ is odd with respect to charge $e$.
This expression  is similar to the induced charge density  in a
pure Coulomb field obtained in \cite{12}. It contains divergence and its renormalization should
be carried out using the obvious physical requirement
the total induced charge to vanish.  Due to the nonzero mass $m$
the renormalization can be performed by usual way in momentum space:
\bb
j_0(\beta)\equiv \rho(\beta) =\int d{\bf r} e^{i{\bf q\cdot r}} j_{0} (r)
=\frac{2e}{\pi} \sum_{l=0}^{\infty} \int\limits^{\infty}_{0} dx
\int\limits^{\infty}_{0} dt \int\limits^{\infty}_{0} dy \frac{\sinh
t}{2b} e^{-y \cosh t} J_{0} (\beta y \sinh t/2b) f(y,t), \nonumber \\
f(y,t)=\frac{xy}{b} \sin(\mu t) (I^{\prime}_{2\gamma_{+}} (y)
+I^{\prime}_{2\gamma_{-}} (y)) - 2a \coth t \cos(\mu t)
(I_{2\gamma_{+}} (y) +I_{2\gamma_{-}} (y)). \label{Imp}
\ee
Here  $\beta =|q|/m$, $x=E/m$, $b=\sqrt{1+x^2}$, $y=2bR/\sinh t$, $R=mr$, $\mu=2ax/b$.

Further calculations with this term for $a<1/2, \alpha\sim 0$ are similar to those described in detail
in \cite{12,ms0} for the vacuum polarization  in a pure Coulomb field in
the subcritical range. We introduce the renormalized induced quantity  in momentum
representation as $n(\beta)=\lim_{\Lambda\to \infty}[\rho(\beta)-\lim_{\beta\to 0}\rho(\beta)]$ with an ultraviolet
cutoff $|E|<\Lambda$ (see, \cite{12,ms0}).  Because the nonzero mass $m$ is the only
 dimensionful parameter in the Green function  the resulting dimensionless function
$n(\beta)$ can depend only on the ratio $\beta=q/m$. Accordingly,
it becomes just a constant in the massless limit $m\to 0$, which is denoted
as $Q=\lim_{m\to 0}n(\beta)$. It is obvious that $Q$ is the induced charge density localized in
the point ${\bf r}=0$ in coordinate space. Therefore the induced charge density in
coordinate space has the form  $Q=Q\delta({\bf r})$. Let us calculate $Q$.

As was shown in  \cite{ms0} the induced charge density (as the series in terms of powers of $a$)
for small $a$ contains divergences only in the coefficients of the $a$ and $a^3$ terms.
We give $n_1(\beta)$ with the $a$ term that reflects the linear
one-loop polarization contribution:
\bb
n_1(\beta) = \frac{2e}{\pi} \sum_{l=0}^{\infty}
\int\limits^{\infty}_{0} dt \left( \int\limits^{\infty}_{0} dy
\int\limits^{\infty}_{0} dx \frac{\sinh t}{2b} e^{-y \cosh t} J_{0}
(\beta y \sinh t/2b) f_1 (y,t) + a \coth t
 (e^{-2 \nu_{+} t} + e^{-2 \nu_{-} t}) \right),   \label{Reg}
\ee
as well as the renormalized induced charge $Q_1$ in the first order of $a$:
\bb
Q_1 = \frac{2ea}{\pi} \sum_{l=0}^{\infty} \int\limits^{\infty}_{0}
dt \left( \int\limits^{\infty}_{0} dy \sinh t \ln (1/y \sinh t) e^{-y
\cosh t}
\times \right. \phantom{mmmmmm}\nonumber \\
\left. \times \left[ yt (I^{\prime}_{2\nu_{+}} (y) +
I^{\prime}_{2\nu_{-}} (y)) - \coth t (I_{2\nu_{+}} (y) +
I_{2\nu_{-}} (y)) \right] + a \coth t
 (e^{-2 \nu_{+} t} + e^{-2 \nu_{-} t}) \right)= \nonumber \\
= \frac{ea}{\pi} \sum_{l=0}^{\infty} \int\limits^{\infty}_{0} dt
\left( \int\limits^{\infty}_{0} dy \ln (1/y \sinh t) \left[ t \sinh t
\frac{d}{dy} (y e^{-y \cosh t} (I_{2\nu_{+}} (y) + I_{2\nu_{-}}
(y))) -
\right. \right. \nonumber \\
\left. \left. - \frac{d}{dt} ( t \cosh t  e^{-y \cosh t})
(I_{2\nu_{+}} (y) + I_{2\nu_{-}} (y))\right] + a \coth t
 (e^{-2 \nu_{+} t} + e^{-2 \nu_{-} t}) \right). \label{Imp2}
\ee
Integrating (\ref{Imp2}), we obtain
\bb Q_1= \frac{2ea}{\pi}
\sum_{l=0}^{\infty} \left( (l+1/2+\alpha) \psi^{\prime}
(l+1/2+\alpha) +(l+1/2-\alpha) \psi^{\prime} (l+1/2 - \alpha) - 2
- \frac{l+1/2}{(l+1/2)^2 - \alpha^2}
\right), \label{Q1}
\ee
where $\psi(z)$ is the logarithmic derivative of Gamma function \cite{GR}.
 For $\alpha \ll 1$,  we find
\bb
Q_1 = ea\pi/4 + ea\pi (2 \ln2 + 1 -\pi^2/4) \alpha^2  \approx ea\pi(0.25 -0.04\alpha^2).
 \label{ApprQ}
\ee
The first term in  Eq. (\ref{ApprQ}) coincides  with result obtained in \cite{12,as11b,bsaso}. We  note that the contribution into $Q_1$ from AB potential arises in the presence of Coulomb field only,  is small and has opposite sign compared with a pure Coulomb one.

We have carried out long calculations and got the total exact induced charge in the subcritical range in the form
\bb
Q = Q_1 + Q_r, \label{Tot}
\ee
where
\bb
Q_r = \frac{2e}{\pi} \sum^{\infty}_{l=0}{\rm Im} \left[ \ln (\Gamma
(\gamma_{+} - ia)\Gamma (\gamma_{-} - ia)) + \frac{1}{2} \ln (
(\gamma_{+} - ia)(\gamma_{-} - ia)) - \right. \nonumber \\
- ((\gamma_{+} -ia) \psi (\gamma_{+} -ia) + (\gamma_{-} -ia) \psi
(\gamma_{-} - ia)) + ia \frac{l+1/2}{(l+1/2)^2 - \alpha^2} -
\nonumber \\
\left. - ia ((l+1/2 +\alpha) \psi^{\prime} (l+1/2 +\alpha) +
(l+1/2 - \alpha) \psi^{\prime} (l+1/2 - \alpha)) \right].
\label{Qr}
\ee
This expression at $\alpha=0$ is in agreement  with result obtained in \cite{12};
the coefficient of the $a^3$ term at $\alpha=0$ was also found in perturbation
theory \cite{bsaso}.
The induced charge $Q$ determined by Eq. (\ref{Tot}) is negative.

It is worth to note that the vacuum charge density
 is induced by the homogeneous background magnetic field in
the massive and massless ${\rm QED}_{2+1}$ \cite{khmam}.

\subsection{Induced charge density in the supercritical range}

In the supercritical range $q>q_c (\gamma=i\sigma)$ the stronger singularity of
the Coulomb potential at the origin has to be regularized, therefore, we need to determine
the self-adjoint Dirac Hamiltonians specified, for example, by self-adjoint boundary conditions (\ref{bouncon}).
Then, we straightforward construct the Green function in the form (\ref{green5}) in which the regular solutions $U_R(r)$,
satisfying (\ref{bouncon}), have to be chosen in the form of
linear combination of the functions $F_R(r)$ and $F_I(r)$.
For this range ($\gamma=i\sigma$),  the above two solutions $F_R(r)$ and $F_I(r)$ become
oscillatory  with the imaginary exponent and  it is convenient to use in this range
the self-adjoint extension parameter  $\theta$ \cite{khlee1,20}, related to $\xi$ by
\bb
\frac{A_R}{\xi A_I} = e^{2i\theta} \left(\frac{2\lambda}{E_0}\right)^{-2i\sigma}
\frac{\nu +a(m+E)/\lambda +is\sigma}{\nu +a(m+E)/\lambda
-is\sigma} \frac{\Gamma(2i\sigma)}{\Gamma(1/2-s/2-aE/\lambda
+i\sigma)} - \frac{\Gamma(-2i\sigma)}{\Gamma(1/2-s/2-aE/\lambda
-i\sigma)},\label{xithet}
\ee
where $\pi\geq \theta\geq 0$ and a positive constant $E_0$ gives an energy scale.

The Green function has a discontinuity, which is solely associated with the appearance
of its singularities situated on a second (unphysical) sheet ${\rm Re}E<0, {\rm Im}E<0$
of the complex plane $E$ at $q>q_c$; these
singularities are determined by complex roots of equation ${\rm W}(E, i\sigma)=0$ and describe
the infinite number of quasistationary (resonant) states with complex ''energies'' $E=|E|e^{i\tau}$.
For massless fermions ($m=0$) and $\sigma\ll 1$  their energy spectrum was found in \cite{20}:
\begin{align}
 E_{k,\theta,s}\equiv {\rm Re}E = E_0 \cos(\tau)
 \exp(-k/2\sigma+\theta/\sigma+\pi\coth\pi a/2a),
\label{energyr}
\end{align}
where $\tau \approx-(1+s)/4a+\mathrm{Im}\psi(ia)+\pi/2$.  Eq. (\ref{energyr}) contains an essential
singularity. These quasi-localized resonances have negative energies, thus they are situated in the hole sector.
For $\sigma\ll 1$ the imaginary part
${\rm Im}E= \tan\tau E_{k,\theta,s}\ll {\rm Re} E$ is very small and, therefore, the resonances are practically stationary states \cite{20}.  For example, for $a=1/2, s=1$  $\tau \approx (1+0.04)\pi$.

Physically, the self-adjoint extension parameter can be
interpreted  in terms of the cutoff radius $R$
of a Coulomb potential. For this, for example, we can compare  Eq. (\ref{energyr}) with the spectrum
of supercritical resonances in the cutoff Coulomb potential \cite{as11b,ggg,gss1}and
 approximately derive $\theta \sim \sigma[c(a)+ \ln E_0R]$, where $c(a)$ does not depend on $R$.
We  note that  the cutoff radius $R$ rather relates to a renormalized critical coupling
that is also characterized by a logarithmic singularity at $mR\ll 1$ in massive case \cite{ggg,khho}

The simplest way to include these resonances in the induced charge density is to carry out the integral
in $E$ from $-\infty$ to $0$ along the path $S$ taking into account the singularities on the second sheet.
After some calculations, we represent the induced charge (electron) density (\ref{cur0}) as
the sum of contributions from the subcritical and supercritical ranges, which have to be treated separately
\bb
j_0(r) = -\frac{e}{2} \int\frac{dE}{2\pi i}{\rm tr}
G_{\nu} (r, r^{\prime}, E) \gamma^0 =-\frac{e}{2} \int\limits_{C}\frac{dE}{2\pi
i}\sum^{+1}_{s=-1} \sum^{+\infty}_{l=-\infty}\frac{f_I(r, \gamma, E) f_R(r, \gamma, E) + g_I(r, \gamma, E)g_R(r, \gamma, E)}{s{\rm W}(E, \gamma)}-\nonumber \\ -\frac{e}{2} \int\limits_{S}\frac{dE}{2\pi
i}\sum_{l,s: \nu<a} \frac{\xi (f_I^2(r, i\sigma, E) + g_I^2(r, i\sigma, E))}{s{\rm W}(E, i\sigma)}
=j_{sub}(r) +j_{sup} (r).\phantom{mmmmmm} \label{Den}
\ee
For the supercritical range  $\gamma=i\sigma$, $0\geq \theta\geq \pi$, the sum in second term $j_{sup}$ is taken  over $l$ of  $a^2>(l+\mu+s/2)^2$.
Then the paths $C, S$ can be deformed to coincide with the imaginary axis $E$.

The first term in Eq. (\ref{Den}) was calculated and explicitly represented in previous subsection.
The second term is convergent and  its contribution to
the induced charge density can be directly evaluated at $m=0$. Having performed simple calculations
we leads $j_{sup}$ to
\bb
j_{sup}(r) =\frac{e}{8\pi^2 r^2}\sum_{l,s: \nu<a}
\frac{s\nu^{s+1}}{\sigma \Gamma(2i\sigma) \Gamma(-2i\sigma)}
\int\limits_{-\infty}^{0} \frac{dE}{E\omega(\sigma)} \Gamma(i\sigma +(1-s)/2 -iaE/|E|)\times \nonumber \\
\times \Gamma(-i\sigma
+(1-s)/2 -iaE/|E|) W_{iaE/|E| +s/2, i\sigma} (2|E|r)W_{iaE/|E| -s/2, i\sigma}
(2|E|r), \label{supcr1}
\ee
where
\bb
\omega(\sigma) = 1- e^{2i\theta} \left(
\frac{2|E|}{E_0}\right)^{-2i\sigma} \frac{\nu+iaE/|E|
+is\sigma}{\nu+iaE/|E| -is\sigma}
\frac{\Gamma(2i\sigma)}{\Gamma(-2i\sigma)}\frac {\Gamma(-i\sigma
+(1-s)/2 -iaE/|E|)} {\Gamma(i\sigma +(1-s)/2 -iaE/|E|)}.
\label{omeg1}
\ee

Rewrite  $(2|E|/E_0)^{-2i\sigma}$ as $\exp(-2i\sigma \ln (|E|/E_0))$. As far as the integrand (\ref{supcr1}) decreases exponentially at $|E|\gg 1/r$ and strongly oscillate at $|E|\to 0$,
the main contribution to the integral over $E$  is given by the region $|E|\sim 1/r$.
So in order to evaluate $j_{sup}$ we replace $|E|$ by $1/r$ in the log-periodic term of the integrand
(\ref{omeg1}) and  obtain
\bb
j_{sup}(r) = -\frac{e}{8\pi^2 r^2}\sum_{l,s: \nu<a}
\frac{s\nu^{s+1}\Gamma(i\sigma
+(1-s)/2 +ia)}{\sigma\omega_-(\sigma) \Gamma(2i\sigma) \Gamma(-2i\sigma)}
\Gamma(-i\sigma +(1-s)/2 +ia) \times \nonumber \\
\times \int\limits_{0}^{\infty} \frac{dE}{E}W_{-ia +s/2,
i\sigma}(2Er)W_{-ia -s/2, i\sigma} (2Er), \phantom{mmmmmmmmmm} \label{supcr2}
\ee
where
\bb
\omega_-(\sigma) = 1- e^{2i\theta +2i\sigma \ln(E_0 r)}
\frac{\nu - ia +is\sigma}{\nu - ia -is\sigma}
\frac{\Gamma(2i\sigma)}{\Gamma(-2i\sigma)}\frac {\Gamma(-i\sigma
+(1-s)/2 + ia)} {\Gamma(i\sigma +(1-s)/2 + ia)}. \label{omega1}
\ee
Because of the complex singularities on the unphysical sheet at $q>q_c$,
the Green function and  $j_{sup}(r)$ are complex though for $\sigma\ll 1$ their imaginary parts
are small.  In terms of the physics the complex Green function probably reflects the lack of
stability of chosen (for constructing Green function) neutral vacuum
for $q>q_c$ (see, also \cite{grrein}).

Now we can integrate  in Eq. (\ref{supcr2}) using formula \cite{GR}
\bb
\int\limits_{0}^{\infty} \frac{dE}{E}W_{-ia +s/2,
i\sigma}(2Er)W_{-ia -s/2, i\sigma} (2Er) = \frac{\pi}{s \sin(2\pi i\sigma)}\times \phantom{mmmmmmmmmm} \nonumber
\\ \times \left[
\frac{1}{\Gamma((1-s)/2 +ia +i\sigma) \Gamma((1+s)/2 +ia
-i\sigma)} -  \frac{1}{\Gamma((1-s)/2
+ia -i\sigma) \Gamma((1+s)/2 +ia +i\sigma)} \right]
 \label{F1}
\ee
and after simple transformations we  finally find the induced charge density in the supercritical range as
\bb
j_{sup}^r(r) = \frac{e}{2\pi^2  r^2}\sum_{l,s: \nu<a}{\rm Re}\frac{\sigma}{\omega_-(\sigma)}. \label{sup4}
\ee

The main effect, arising at supercritical regime, is that the induced
vacuum polarization for noninteracting massless fermions has a power law form ($\sim c/r^2$)  whose
coefficient is log-periodic functions with respect
to the distance from the origin. In the subcritical regime the induced vacuum
charge  is localized at origin and exhibits no long range tail.
As an example, we consider Eq. (\ref{sup4}) for $1/2-\alpha<a<3/2+\alpha, 1/2\gg \alpha>0$, when just the lowest $l+n, s$ channels are supercritical, and find
\bb
j_{sup}^r(r) = \frac{e}{2\pi^2  r^2}\sum_{\sigma=\sigma_{\pm}}\sigma{\rm Re}\frac{2-Az e^{2i\theta +2i\sigma \ln(E_0r)}}{1-Az e^{2i\theta +2i\sigma \ln(E_0r)}+A^2[(a-\sigma)/(a+\sigma)]e^{4i\theta +4i\sigma \ln(E_0r)}}, \label{2sup}
\ee
where
\bb
 A=\frac{\Gamma(2i\sigma)\Gamma(-i\sigma+ia)}{\Gamma(-2i\sigma)\Gamma(i\sigma+ia)},\quad
 z=2\frac{a-\sigma}{a},\quad \nu_{\pm}=1/2\pm \alpha, \sigma_{\pm}=a^2-\nu_{\pm}^2.
\label{noti2}
\ee
The induced charge density (\ref{2sup}) resembles  the local density of states which also exhibits resonances
at the negative energies \cite{20}.

Using the known representation
$$
{\rm Arg}\Gamma(x+iy)=y\left[-{\cal C}+\sum\limits_{n=1}^{\infty}\left(\frac1n-\frac1y\arctan\frac{y}{x+n-1}\right)\right],
$$
where ${\cal C}=0.57721$ is Euler's constant, we finally obtain the induced charge density (\ref{2sup}) in the form
\bb
j_{sup}^r(r) = \frac{e}{2\pi^2  r^2}\sum_{\sigma=\sigma_{\pm}}\sigma{\rm Re}\frac{2-|A|ze^{2i\theta +2i\sigma \ln(E_0 r)+i\psi}}{1-|A|ze^{2i\theta +2i\sigma \ln(E_0 r)+i\psi}+|A|^2[(a-\sigma)/(a+\sigma)]e^{4i\theta +4i\sigma \ln(E_0 r)+2i\psi}}. \label{fsup}
\ee
Here
\bb
\psi \equiv {\rm Arg} A = -\pi-2{\cal C}\sigma +\sum\limits_{n=1}^{\infty}\left(\frac{2\sigma}{n}-2\arctan\frac{2\sigma}{n}
+\arctan\frac{2n\sigma}{n^2+\nu^2}\right).
\label{not5}
\ee
For small $\sigma\ll 1$, Eq. (\ref{fsup}) takes the simplest form
\bb
j_{sup}^r(r) = \frac{e (\sigma_++\sigma_-)}{2\pi^2  r^2},\quad \sigma_{\pm}=\sqrt{a^2-(1/2 \pm \alpha)^2}. \label{1fsup}
\ee
Here $``\pm"$ sign before $\alpha$ for fixed sign of $\alpha$, in fact, corresponds to the spin projection sign.
One sees that $j_{sup}^r(r)$ is odd with respect to the fermion charge $e$ and even with respect to $\alpha$. 
The contribution into the induced charge density due to the AB potential  has opposite sign compared with a pure Coulomb one.
It is of importance that the induced charge density $j_{sup}^r(r)$ (\ref{1fsup})  at $\sigma\ll 1$ does not contain at all the self-adjoint extension parameter $\theta$. From the physical point of view, increased $a$ near the point
($\gamma=0$) the transition will occur from the subcritical range to the supercritical one,
which can be symbolically characterized by $\gamma\to \sigma$ and then a small change
in $\sigma$ such that $q>q_c$ leads to a sudden change in the character of a physical phenomenon due
to emerging of infinitely many resonances with negative energies.
Put another way the character of a physical phenomenon itself must be due only to physical  (but not mathematical)  reasons.
We also note that the expression $j_{sup}^r(r)$ (at $\alpha=0$) is in agreement  with results
obtained in \cite{as11b} for the problem of vacuum polarization of supercritical
impurities in graphene by means of scattering phase analysis.

\section{Concluding remarks}

In this paper we have studied the vacuum polarization of charged massless fermions  in
 Coulomb and AB potentials in 2+1 dimensions.
In particular, we have calculated the induced  charge density using the Green's functions of
the Dirac equation with the Coulomb and AB  potentials. In subcritical  regime
the induced vacuum charge $Q$ is localized at the origin  and
has a screening sign, leading to a decrease of the effective Coulomb
charge; the contribution into $Q$ due to the AB potential  is  small and has opposite sign compared to the Coulomb one.
In the supercritical regime the induced vacuum charge, like the subcritical contributions,  has a screening
sign but it has a power law form, causing a modification of Coulomb's law at large distances; the contribution into
the induced vacuum charge  due to the AB potential  has opposite sign compared to the Coulomb one.

Because a single electron dynamics in graphene is described by a massless
two-component Dirac equation  we hope that our results can
be applied in graphene with charged impurities.
 Furthermore, while the electron-electron interaction has been neglected,
results of present paper may be useful to develop further
insight into the screening of the Coulomb impurity with taking into consideration of
the electron spin and  electron-electron interaction.
To approach this problem one can write the self-consistent
renormalization group equations in the Hartree approximation
in the subcritical range in the same spirit as in \cite{12}  and
within the Thomas-Fermi method for the supercritical range.
We  shall defer the self-consistent renormalization group analysis
 to a future work.

\end{document}